\documentclass[twocolumn,showpacs,preprintnumbers,amsmath,amssymb]{revtex4}
\usepackage{graphicx}
\usepackage{dcolumn}
\usepackage{bm}
\begin{document}

\title{The Fractional Quantum Hall States of Dirac Electrons in Graphene}
\author{Vadim M. Apalkov} 
\affiliation{Department of Physics and Astronomy, Georgia State University, 
Atlanta, Georgia 30303, USA}
\author{Tapash Chakraborty$^\ddag$}
\affiliation{Department of Physics and Astronomy,
University of Manitoba, Winnipeg, Canada R3T 2N2}

\date{\today}
\begin{abstract}
We have investigated the fractional quantum Hall states of 
Dirac electrons in a graphene layer in different Landau 
levels. The relativistic nature of the energy dispersion relation 
of electrons in graphene significantly modifies the inter-electron 
interactions. This results in a specific dependence of the ground 
state energy and the energy gaps for electrons on the Landau-level 
index. For the valley-polarized states, i.e. at $\nu =1/m$, $m$ 
being an odd integer, the energy gaps have the largest values in 
the $n=1$ Landau level. For the valley-unpolarized states, e.g., 
for the $2/3$ state, the energy gaps are suppressed for 
$n=1$ as compared to those at $n=0$. For both $n=1$ and $n=0$ 
the ground state of the $2/3$ system is fully valley-unpolarized. 

\end{abstract}
\maketitle

A two-dimensional electron system in a single layer of graphite (graphene)
is known to exhibit many remarkable properties. From the band structure 
studies \cite{wallace} it was established early on that, to a good 
approximation, the energy dispersion of electrons in graphene is {\it 
linear} near the points at the corners of the Brillouin zone where the 
valence band and the conduction band meet. As a consequence, the 
low-energy excitations follow the Dirac-Weyl equations for massless 
{\it relativistic} particles \cite{ando}. In an external magnetic field, 
the electron system also shows unique properties that are different 
from those of the standard {\it non-relativistic} electron systems 
\cite{ando,mcclure,landau_levels}. Recent experimental demonstration of 
some of those properties, in particular, the discovery of the integer
quantum Hall effect \cite{qhe_expt,high_field} that was predicted in 
earlier theoretical works \cite{qhe_theory} has caused intense interest 
in the electronic properties of the Dirac electrons in graphene 
\cite{phys_today}. However, effects of electron correlations in this 
system have not been reported yet. In this paper, we report on the nature 
of the fractional quantum Hall states of Dirac electrons in graphene.

A unit cell of the two-dimensional (2D) graphene honeycomb lattice contains 
two carbon atoms, say $A$ and $B$. The dynamics of electrons in graphene 
is described by a tight-binding Hamiltonian with the nearest-neighbor 
hopping. In the continuum limit this Hamiltonian generates the band structure 
with two $\pi $ bands and the Fermi levels are located at two inequivalent 
points, $K = (2\pi /a) (1/3,1/\sqrt{3})$ and $K^{\prime } = (2\pi /a) 
(2/3,0)$, of the first Brillouin zone, where $a=0.246$ nm is the lattice 
constant. Near the points $K$ and $K^{\prime}$ the electrons have a linear 
Dirac-Weyl (``relativistic'') type dispersion relation \cite{wallace,ando}. 
Finally, in the continuum limit the electron wave function is described 
by the 8-component spinor, $\Psi_{s,k,\alpha}$, where $s=\pm 1/2$ is the 
spin index, $k = K, K^{\prime}$ is the valley index, and $\alpha = A, B$ 
is the sublattice index. Without the spin-orbit interaction 
\cite{kane05,wang06,sinitsym05} the Hamiltonian is described by two 
$4\times 4$ matrices for each component of the electron spin. In the 
presence of a magnetic field perpendicular to the graphene plane the 
Hamiltonian matrix has the form
\begin{equation}
{\cal H} = \frac{\gamma}{\hbar } \left( 
\begin{array}{cccc}
    0 & \pi_x - i \pi_y & 0 & 0   \\
    \pi_x + i \pi_y & 0 & 0 & 0 \\
    0 & 0& 0& \pi_x + i \pi_y   \\
    0 & 0&  \pi_x - i \pi_y & 0 
\end{array} 
\right) ,
\label{H}
\end{equation}
where $\vec{\pi } = \vec{p} + e\vec{A}/c$, $\vec{p}$ is the
two-dimensional momentum, $\vec{A}$ is the vector potential, and $\gamma $ 
is the band parameter. The ordering used for the basis states in the
non-interacting Hamiltonian is $(K,A;K,B;K^{\prime },A;K^{\prime },B)$. 
The eigenfunctions of the Hamiltonian are specified by the Landau-level 
index $n=0,\pm 1, \ \pm 2, \ldots$ and the intra-Landau level index $m$ 
that is gauge dependent. Each Landau level is four-fold degenerate due 
to the spin and valley degrees of freedom. The corresponding wave functions 
for an electron in the two valleys $K$ and $K^{\prime}$ are described by
\begin{equation}
\Psi_{K,n} = C_n
\left( \begin{array}{c}
 {\rm sgn} (n) i^{|n|-1} \phi_{|n|-1} \\
    i^{|n|} \phi _{|n|} \\
    0  \\
    0
\end{array}  
 \right),
\label{f1}
\end{equation}
\begin{equation}
\Psi_{K^{\prime},n} = C_n
\left( \begin{array}{c}
   0 \\
   0\\
    i^{|n|} \phi _{|n|} \\
 {\rm sgn} (n) i^{|n|-1} \phi_{|n|-1} 
\end{array}  
 \right),
\label{f2}
\end{equation}
where $C_n = 1 $ for $n=0$ and $C_n = 1/\sqrt{2}$ for $n\neq 0$.
Here $\phi _n$ is the standard Landau wave function for a
particle with non-relativistic parabolic dispersion relation in
the $n$-th Landau level. From Eqs.~(\ref{f1})-(\ref{f2}) it is
clear that a specific feature of the relativistic dispersion law 
is the mixing of the non-relativistic Landau levels. This mixture 
is present only for $n\neq 0$ and strongly modifies the inter-electron
interaction within a single Landau level.

In what follows, we study the partially occupied Landau levels with
fractional filling factors. Partial occupation of the Landau levels is
realized by doping of the graphene layer. Experimentally, different filling 
factors of the Landau levels are achieved by varying the applied magnetic field
at a fixed electron concentration. In this case the ground state of
the system and the excitation spectrum are fully determined by
the inter-electron interactions. For the non-relativistic case this
results in the incompressible fractional quantum Hall effect (FQHE) states 
at the fractional filling factors \cite{fqhe_book,fqhe_review}. Properties 
of these states are completely described by Haldane's pseudopotentials, 
$V_m$ \cite{haldane}, which are the energies of two electrons with 
relative angular momentum $m$. The pseudopotentials for the $n$-th 
Landau level can be presented as
\begin{equation}
V_m^{(n)} = \int _0^{\infty } \frac{dq}{2\pi} q V(q) 
\left[F_n(q) \right]^2 L_m (q^2) 
 e^{-q^2}, 
\label{Vm}
\end{equation}
where $L_m(x)$ are the Laguerre polynomials, $V(q) = 2\pi e^2/(\kappa l q)$ 
is the Coulomb interaction in the momentum space, $\kappa$ is the 
dielectric constant, $l$ is the magnetic length, and $F_n(q)$ is the form 
factor corresponding to the $n$-th Landau level. For relativistic electrons 
the form factor is given by the expression \cite{goerbig06}
\begin{eqnarray}
& & F_0(q) = L_0\left( \frac{q^2}{2} \right)  \label{f0}  \\
& & F_{n \not=0}(q) = \frac{1}{2} \left[ L_n \left(\frac{q^2}{2} \right) 
  +  L_{n-1} \left( \frac{q^2}{2} \right)    \right], 
\label{fn}
\end{eqnarray}
while for the non-relativistic particles the form factors in Eq.~(\ref{Vm})
are $F_n (q) = L_n\left( q^2/2\right)$. This means that the inter-electron 
interactions for the relativistic and non-relativistic electrons are the 
same for $n=0$ and are different for $n>0$ \cite{goerbig06}. In what follows, 
all energies are expressed in units of the Coulomb energy 
$\varepsilon_c=e^2/\kappa l$.

\begin{figure}
\begin{center}\includegraphics[width=9cm]{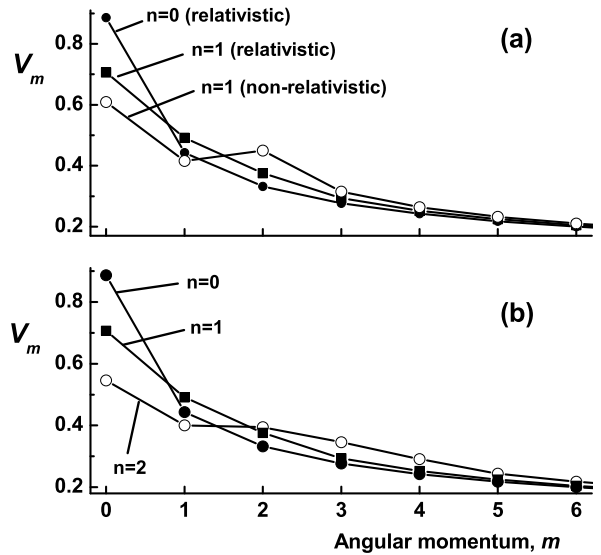}\end{center}
\vspace*{-1cm}
\caption{The pseudopotentials [Eq.~(\ref{Vm})] as a function of the relative 
angular momentum (a) for relativistic and for non-relativistic 2D electrons in 
the first two Landau levels, and (b) for relativistic electrons in various 
Landau levels. 
}
\label{figone}
\end{figure}

In Fig.~1 we compare the pseudopotentials calculated from
Eq.~(\ref{Vm}) for the relativistic and the non-relativistic cases.
For the relativistic electrons [Fig.~1 (a)] we notice a clear
suppression of the pseudopotential for $n=1$ as compared to that at $n=0$,
only for $m=0$, i.e. when both electrons are at the same spatial
point. For all other $m$, we have the inequality $V_m^{(1)} > V_m^{(0)}$.
This is different from the non-relativistic case where the
pseudopotential is suppressed also for $m=1$, i.e.  $V_1^{(1)} < V_1^{(0)}$. 
We also see in Fig.~1 (a) that although the relativistic wave functions at 
$n=1$ is the ``mixture'' of the $n=0$ and $n=1$ non-relativistic
wave functions, the relativistic pseudopotential is not the average
of the corresponding non-relativistic pseudopotentials. This is clearly
seen for $m=1$ where the relativistic pseudopotential at $n=1$  
is larger than the non-relativistic one for both $n=0$ and $n=1$.
In Fig.~1 (b) the relativistic pseudopotentials are shown for different
Landau levels. Here the special case is $m=1$ where the dependence of 
the pseudopotential on the Landau-level index is non-monotonic, viz., the 
pseudopotential has the maximum value at $n=1$. At all the other $m$ values
the trend is the same as for the non-relativistic case, i.e., for $m=0$ the 
pseudopotential decreases with increasing $n$, while at $m>1$ the 
pseudopotentials increase with $n$.

With the pseudopotentials for Dirac electrons at hand, we now evaluate
the energy spectra of the many-electron states at fractional fillings
of the Landau level. The calculations have been done in the spherical
geometry \cite{haldane} with the pseudopotentials given by Eq.~(\ref{Vm}). 
The radius of the sphere $R$ is related to $2S$ of magnetic fluxes 
through the sphere in units of the flux quanta as $R = \sqrt{S} l$. The 
single-electron states are characterized by the angular momentum $S$, 
and its $z$ component $S_z$. For a given number of electrons $N$, the 
parameter $S$ determines the filling factor of the Landau level. Due to 
the spherical symmetry of the problem, the many-particle states are 
described by the total angular momentum $L$, and its $z$ component, 
while the energy depends only on $L$. At first we study the system with
the fractional filling factor $\nu = 1/3$, which corresponds to the
$1/3$ FQHE. In the spherical geometry, the $1/3$-FQHE state is realized 
at $S = (3/2)(N-1)$. If the electron system is fully spin and valley
polarized then we should expect the ground state to be in the Laughlin
state \cite{laughlin} which is separated from the excited states
by a finite gap. We calculated the energy spectra of a finite-size
system by finding the lowest eigenvalues and eigenvectors of
the interaction Hamiltonian matrix \cite{fano}. In these calculations 
we take into account the interaction between all the electrons of the partially
occupied Landau levels. We have also addressed the question of polarization 
of the many-particle state.  We assume that at a high magnetic field the 
system is always spin-polarized. In this case the system can be
either valley-polarized or valley-unpolarized. Similar to the spin 
polarization of the standard FQHE states the valley polarization of the
graphene system depends on the inter-valley asymmetry (same as the Zeeman 
splitting for the non-relativistic system) and filling factor of the Landau
level. Namely, if the inter-valley splitting due to the inter-valley asymmetry 
is large then the system is always valley-polarized, while otherwise
the valley polarization of the system depends on the filling factor of the
Landau level.

\begin{figure}
\begin{center}\includegraphics[width=9cm]{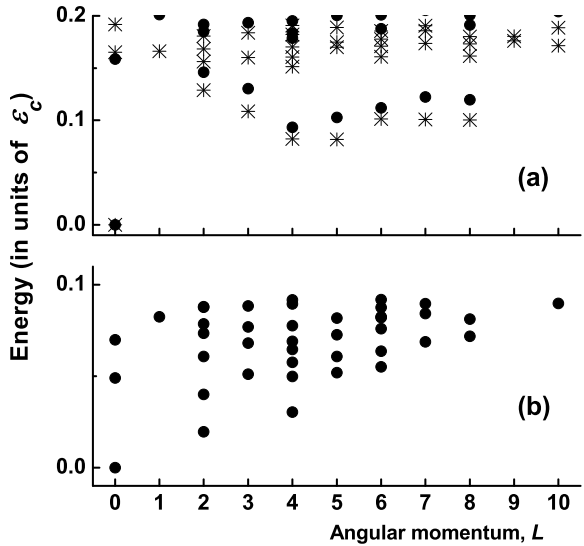}\end{center}
\vspace*{-1cm}
\caption{Energy spectra of the eight-electron $\nu =1/3$-FQHE
system, shown for different Landau levels: (a) $n=0$
(stars) and $n=1$ (dots), and (b) $n=2$. The system is fully spin 
and valley polarized. The flux quanta is $2S = 21$. 
}
\label{figtwo}
\end{figure}

In Fig.~2 we show the calculated energy spectra for the $1/3$-FQHE
state at different Landau levels. Here, the $1/3$-FQHE state
at the $n$-th Landau level is defined as the state corresponding to
the $1/3$ filling factor (single valley, single spin) of the $n$-th 
Landau level, while all the lower energy Landau levels are completely 
occupied. Since the relativistic pseudopotential $V_m^{(0)}$ for $n=0$ 
is the same as for the non-relativistic one, the $1/3$ state and the 
corresponding energy gap (in units of $\varepsilon_c$) will be the same 
in both cases. The deviation from the non-relativistic system occurs 
only at higher Landau levels. In Fig.~2 (a) the energy gap of the 
$1/3$-state at $n=1$ is noticeably enhanced compared to that at $n=0$. 
This is a direct manifestation of the specific dependence of pseudopotenials 
$V_m^{(n)}$ on the Landau-level index. Due to the asymmetry of the 
electron wave functions, the spectra of the $1/3$-FQHE state is mainly 
determined by the relative value of $V_1^{(n)}$ and $V_3^{(n)}$ 
pseudopotentials, which have the highest value at $n=1$. The energy 
spectra of the $1/3$-FQHE state at $n=2$, shown in Fig.~2 (b), demonstrate 
a strong suppession of the gap when compared to the $n=1$ and $n=0$ FQHE 
states. We therefore conclude that {\it the $1/3$-FQHE state in graphene
is most stable at $n=1$}. Hence the inter-electron interaction effects 
are more pronounced at $n=1$. Interestingly, this tendency is just the 
opposite to that of the non-relativistic system, where the excitation gap 
decreases monotonically with incresing Landau-level index \cite{fqhe_book}. 

The results in Fig.~2 correspond to a completely spin and valley polarized 
system. This polarization is achieved at a high magnetic field due to the 
Zeeman splitting and the valley asymmetry. The inter-valley asymmetry has 
two sources: the first one is due to interaction-induced ``backscattering'' 
between different valleys \cite{goerbig06} while the second one is due 
to the asymmetry in the lattice-scale interactions within the two 
sublattices of graphene \cite{fisher06}. Since the positions of the
electrons in two sublattices are shifted, the interaction between the 
electrons in the different sublattices is weaker than the interaction 
between the electrons in the same sublattice. Both effects vary as 
$(a/l)$, so they become more relevant at higher magnetic fields or at 
a smaller magnetic length.

The same picture holds for the other FQHE states of the type $1/m$, i.e., 
the state is most stable at $n=1$. The new aspects of the interaction 
physics arise at other filling factors as well, when the lowest energy 
states are spin-unpolarized for the non-relativistic single-valley 
systems. The simplest example is $\nu =2/3$. In this case the ground 
state of the non-relativistic electrons is spin-unpolarized at a
small Zeeman splitting. The transition from the spin-polarized to the 
spin-unpolarized ground states in a tilted magnetic field is well
established both experimentally and theoretically for non-relativistic
electrons \cite{fqhe_review}.

\begin{figure}
\begin{center}\includegraphics[width=9cm]{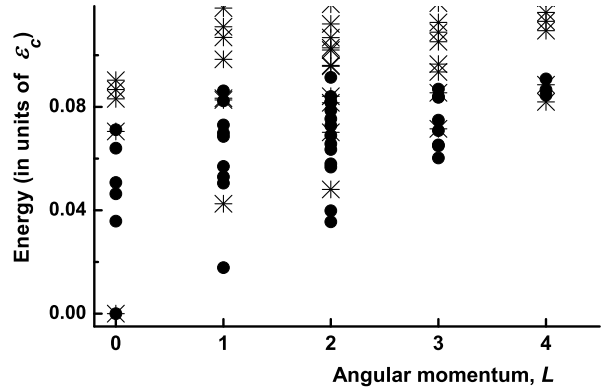}\end{center}
\vspace*{-1cm}
\caption{Energy spectra of the eight-electron $\nu =2/3$-FQHE
system, shown for different Landau levels: $n=0$ (stars) and
$n=1$ (dots). The system is valley-unpolarized and
fully spin-polarized. The flux quanta is $2S = 11$.
}
\label{figthree}
\end{figure}

Here we address the problem of the polarized and unpolarized states 
in the graphene system. We assume that the Zeeman splitting is large 
enough so that all the states are spin-polarized. At the same time 
the inter-valley asymmetry is small and electrons can occupy both the 
valleys. In this case the valley-polarized and valley-upolarized 
$2/3$-FQHE states become relevant. For a valley-polarized system 
the excitation spectra and the ground state properties of the
$2/3$-state are the same as those for $\nu=1/3$ due to the particle-hole 
symmetry. Similar to the $1/3$ case we obtain numerically an enhancement 
of the excitation gap at $n=1$ as compared to that at 
$n=0$. The more complicated situation occurs for the valley-unpolarized 
system. First we compare, just as for the non-relativistic 
system \cite{xie89}, the energy of the ground states of the polarized and 
the unpolarized systems. In the spherical geometry the polarized 
and the unpolarized states are realized for different sizes of the 
sphere, i.e., for different flux quanta through the sphere. For the
$2/3$-FQHE system the polarized state occurs at $2S = 3N/2$, while 
the unpolarized one is at $2S = 3N/2-1$. Due to different size of the 
sphere in these two systems the finite-size corrections to the magnetic 
length should be introduced, $l^{\prime} = (\nu
2S/N)^{1/2}l$ \cite{xie89,morf86}. 

We have calculated the ground state energies for the valley-polarized 
and the valley-unpolarized graphene system in the $n=1$ and the $n=0$ 
Landau levels for a eight-electron system in the spherical geometry. 
We found that for both $n=0$ and $n=1$ the valley-unpolarized state has 
the lower energy. The ground state energy per particle in the unpolarized 
system is lower than that for the polarized system by $0.073 \varepsilon_c$ 
for $n=0$ and by $0.053 \varepsilon_c$ for $n=1$. Therefore, the 
valley-unpolarized state is more favorable for $n=0$. Here the gap 
between the polarized and the unpolarized states is suppressed for $n=1$ 
as compared to the $n=0$ case. This is opposite to the completely 
polarized system (Fig.~2), where the effects of interaction is the 
strongest for $n=1$. Suppression of the interaction effects 
in an unpolarized system for $n=1$ is also illustrated in Fig.~3, where the 
excitation spectra of the valley-unpolarized system is shown for $n=0$ and 
$n=1$. A strong suppression of the excitation gaps for $n=1$ is
clearly visible here. The origin of this suppression can be understood from 
the dependence of the pseudopotentials $V_m^{(n)}$ on the relative angular 
momentum $m$ in different Landau levels. Due to the Pauli exclusion 
principle the energetic properties of the polarized state is determined 
only by the pseudopotentials with odd angular momenta, $m=1,3,5,\ldots$. 
These pseudopotentials have the largest values for $n=1$, 
which results in the strongest interaction effects for $n=1$, 
in a polarized system. For an unpolarized system the properties of 
the ground and excited states depend on all the pseudopotentials. Since 
$V_m^{(n)}$ at $m=0$ is strongly suppressed for $n=1$
as compared to the $n=0$ case, we expect a suppression of 
interaction effects in an unpolarized system in the $n=1$ Landau level. 

In conclusion,  the relativistic nature of the energy dispersion of 
electrons in the graphene plane modifies the inter-electron interactions 
significantly. This results in a unique dependence of the ground state energy 
and the energy gaps of the graphene systems on the Landau-level index. For 
the valley-polarized states, i.e. at $\nu =1/m$, the FQHE gaps have the 
largest values for $n=1$. Based on these studies we conclude 
that the FQHE at $\nu =1/m$ should be observed experimentally for both  
$n=0$ and $n=1$, perhaps in a higher mobility system. 
For the valley-unpolarized states, e.g., for the $2/3$-FQHE state, the 
energy gaps are suppressed at $n=1$ as compared to that for the $n=0$ level. 
For both $n=1$ and $n=0$ the ground state of the $2/3$-FQHE system is fully 
unpolarized. The inter-valley asymmetry will result in transitions between 
the valley-polarized and the valley-unpolarized states. 

The work has been supported by the Canada Research Chair Program 
and a Canadian Foundation for Innovation Grant.

\end{document}